\documentclass{article}
\usepackage{dcase2022_techrep,amsmath,graphicx,url,times,booktabs, tabularx, stfloats}

\title{Frequency Dependent Sound Event Detection for DCASE 2022 Challenge Task 4}

\name{Hyeonuk Nam, Seong-Hu Kim, Deokki Min, Byeong-Yun Ko, Seung-Deok Choi, Yong-Hwa Park \thanks{This work was supported by “Human Resources Program in Energy Technology” of the Korea Institute of Energy Technology Evaluation and Planning (KETEP), granted financial resource from the Ministry of Trade, Industry \& Energy, Republic of Korea. (No. 20204030200050)}}
\address{Korea Advanced Institute of Science and Technology\\
Department of Mechanical Engineering, 291 Daehak-ro,\\
Yuseong-gu, Daejeon 34141, South Korea\\
\{frednam, seonghu.kim, minducky, b.y.ko, haroldchoi6, yhpark\}@kaist.ac.kr}

\begin{document}

\ninept
\maketitle

\begin{sloppy}

\begin{abstract}
While many deep learning methods on other domains have been applied to sound event detection (SED), differences between original domains of the methods and SED have not been appropriately considered so far. As SED uses audio data with two dimensions (time and frequency) for input, thorough comprehension on these two dimensions is essential for application of methods from other domains on SED. Previous works proved that methods those address on frequency dimension are especially powerful in SED. By applying FilterAugment and frequency dynamic convolution those are frequency dependent methods proposed to enhance SED performance, our submitted models achieved best PSDS$_{1}$ of 0.4704 and best PSDS$_{2}$ of 0.8224.
\end{abstract}

\begin{keywords}
Sound Event Detection, FilterAugment, Frequency Dynamic Convolution
\end{keywords}

\section{Introduction}
Sound event detection (SED) which aims to classify desired sound event classes and their time localization (onset and offset) in a given audio signal has been rapidly growing with advancement of deep learning (DL) methods \cite{casse, DCASEtask4, dcase2022task4website, sedmetrics, mytechreport, filtaug, FDY}. As 1D audio data with time dimension is usually expanded into 2D data with time and frequency dimension for audio signal processing, 2D time-frequency audio data are usually used for DL based SED by treating 2D audio data as 2D image data and applying DL methods for image data \cite{filtaug, FDY, coughcam}. Although DL methods for 2D image data showed powerful performance on their own domain, they have inherent inconsistency on SED which arises from the difference between 2D image data and 2D audio data. While 2D image data consists of two same dimensions representing the same physical quantity (location), 2D audio data consists of two different dimensions representing different physical quantity (time and frequency). Considering that time is somewhat similar to location as they both are translation equivariant (certain pattern is still the same entity when it is moved along location or time dimensions) while frequency is not translation equivariant because each frequency value represents different characteristics from the others, frequency is the dimension to be thoroughly considered in SED \cite{filtaug, FDY}. In this work, we especially apply methods that address such difference between 2D image data and 2D audio data for SED by addressing issues of frequency dimension in 2D audio data. SED models illustrated in this report could be trained using code available in GitHub\footnote{https://github.com/frednam93/FDY-SED}. This repository includes both FilterAugment and frequency dynamic convolution.

\section{Methods}
\subsection{FilterAugment}
FilterAugment is proposed to generalize SED model to various acoustic environments, to make SED model more like human who can classify sound events from different acoustic environments \cite{filtaug}. By randomly dividing frequency ranges into several frequency bands and randomly applying weights on the frequency bands of a Mel spectrogram, FilterAugment could approximately simulate acoustic environments that results in different frequency weights on different frequency bands. Although resulting Mel spectrogram might sound unnatural, it is simple to use and effective on SED as shown in \cite{filtaug}. FilterAugment could emphasize different frequency bands of the same data every epoch, thus it helps to train SED model to recognize time-frequency patterns from wider frequency ranges. Without FilterAugment, SED model might be trained to recognize patterns from most distinct time-frequency patterns instead. Code for SED with FilterAugment is available in GitHub\footnote{https://github.com/frednam93/FilterAugSED}.

There are two types of FilterAugment: step and linear type. Step type FilterAugment applies constant weights over each frequency bands and the weights change abruptly across the boundary of frequency bands, while linear type FilterAugment applies continuous weights over frequency bands by assigning weights on frequency boundaries and then linearly interpolating weights between the boundaries. In this work, these different types of FilterAugment ensemble averaged to increase variety of SED model capacity thus enhance performance of ensemble aveaged model.

Hyperparameter setting used in this work is as follows: step type FilterAugment with dB range = (-4.5, 6), band number range = (2, 5) and minimum bandwidth = 4, linear type FilterAugment with dB range = (-6, 4.5), band number range = (3, 6) and minimum bandwidth = 7.

\subsection{Frequency Dynamic Convolution}
Frequency dynamic convolution is proposed to weaken translation equivariance of 2D CNN on frequency axis and to improve CNN kernel's adaptability to the input at the same time \cite{FDY}. It is inspired by temporal dynamic models that applied dynamic convolution proposed for image recognition into speaker verification \cite{acnn, tdycnn, DTDY}. As different frequency regions exhibit different frequency patterns of sound events, different convolution kernels should be used on different frequency regions. Thus, frequency dynamic convolution applies convolution kernel that dynamically adapts to each frequency bin of the convolution input. By applying adaptive kernels that differs on each frequency bin, translation equivariance is weakened in frequency dynamic convolution and it helps recognizing more complex time-frequency pattern for SED as shown in analysis of class-wise comparison \cite{FDY}. In this work, 4 basis kernels and temperature of 45 are used for frequency dynamic convolution.

\subsection{Implementation Details}
The code used to train SED model submitted for this participation could be found in Github repository mentioned in introduction. It is derived from DCASE 2021 Challenge Task 4 baseline \cite{DCASEtask4, dcase2022task4website}. Input audio data with length of 10 seconds and sampling rate of 16 kHz are used. They are converted to log Mel spectrogram with number of FFT, hop length and number of Mel bins as 2048, 256 and 128 respectively. Each batch of log Mel spectrograms are normalized to be between 0 and 1 on batch and time dimensions. For data augmentation, frame shift, mixup \cite{mixup}, time masking \cite{specaug} and FilterAugment \cite{filtaug} are used. To utilize unlabeled dataset, mean teacher method is used \cite{meanteacher}.

The model has CRNN architecture composed of 7 CNN layers and 2 BiGRU layers, then frame-wise fully connected (FC) layer makes strong prediction of the input \cite{crnn}. Each CNN layer is composed of convolution module, batch normalization, context gating, dropout, and then 2D average pooling. The first CNN layer uses normal 2D convolution and the rest six CNN layers use frequency dynamic convolution \cite{FDY}. After 2 biGRU layers, frame-wise FC layer outputs strong prediction while other FC layer followed by Softmax extracts attention weights to apply on strong prediction to result in weak prediction. On the output, weak prediction masking or weak SED \cite{mytechreport} is applied and then applied by median filter. Different length of median filter is applied on each sound event class: 5 for alarm/bell ringing, cat, dish, dog and speech and 11, 67, 61, 49, 17 for blender, electric shaver/toothbrush, frying, running water and vacuum cleaner respectively.

Evaluation metrics used in this report are polyphonic sound detection score (PSDS) \cite{PSDS} and macro event-based F1 score \cite{sedmetrics}. PSDS$_{1}$ and PSDS$_{2}$ are the metrics used in DCASE 2022 challenge Task4 \cite{dcase2022task4website}, which favors SED system that predicts accurate timestamp and SED system that does not produce cross triggers respectively.

\begin{table*}
\caption{Training settings with their best PSDS$_{1}$, PSDS$_{2}$ scores and collar-based F1 score.}
\centering
\setlength{\tabcolsep}{8.5pt}
\begin{tabular}{l|lll|lll}
\hline
\textbf{Setting Index} & \textbf{Seed} & \textbf{FilterAugment Type} & \textbf{Attention Dimension} & \textbf{PSDS$_{1}$} & \textbf{PSDS$_{2}$} & \textbf{CB-F1} \\ \hline
1 & 21  & step   & class  & 0.4446  & 0.6780  & 0.536  \\
2 & 42  & step   & class  & 0.4554  & 0.6719  & 0.536  \\
3 & 42  & linear & class  & 0.4510  & 0.6702  & 0.536  \\
4 & 42  & step   & time   & 0.4548  & 0.6743  & 0.533  \\

\hline
\end{tabular}
\end{table*}

\subsection{Ensemble Averaging}
Table 1 shows four settings used in this work with their best PSDS scores and collar-based F1 score (CB-F1) of single SED model in each setting. For each setting, 48 training runs are separately done resulting in 96 models (48 student models and 48 teacher models). The settings differ by seed, FilterAugment type, and attention type for weak prediction pooling. The seed is used are 21 and 42. FilterAugment types used are step and linear as illustrated in 2.1. Attention types are class and time, which means the dimension on which Softmax applied to obtain attention weights. Using different settings resulted in more diverse SED models thus resulted in better ensemble averaged models.

\section{Results}
Table 2 shows performance of submitted models which are ensemble averaged model of 4 settings in Table 1. We chose best 31 student and teacher models on PSDS$_{1}$ for submission 1, best 12 student models on PSDS$_{1}$ for submission 2,  best 53 student models on PSDS$_{2}$ for submission 3 and best 150 student and teacher models on PSDS$_{2}$ for submission 4. For submission 1 and 2, weak prediction masking is applied and for submission 3 and 4, weak SED is applied \cite{mytechreport}. As a result, the best PSDS$_{1}$ score is .0.4704 and the best PSDS$_{2}$ score is 0.8224.

\begin{table}
\caption{Performance of submissions with number of models ensemble averaged.}
\centering
\setlength{\tabcolsep}{8.5pt}
\begin{tabular}{ll|lll}
\hline
\textbf{Submit Index} & \textbf{\# Models} & \textbf{PSDS$_{1}$} & \textbf{PSDS$_{2}$} & \textbf{CB-F1} \\ \hline
1 & 31  & \textbf{0.4704}  & 0.6866          & 0.543  \\
2 & 12  & 0.4703           & 0.7002          & 0.541  \\
3 & 53  & 0.0606           & \textbf{0.8224} & 0.199  \\
4 & 150 & 0.0584           & 0.8195          & 0.536  \\

\hline
\end{tabular}
\end{table}

\bibliographystyle{IEEEtran}
\bibliography{reference.bib}
%
%
%
%
%
%
%
%
%

\end{sloppy}
\end{document}